\documentclass[journal]{IEEEtran}
\usepackage{cite}
\usepackage{subfigure}
\usepackage{color}
\usepackage{graphicx}
\usepackage{dcolumn}
\usepackage{bm}
\usepackage{physics}
\usepackage{blindtext}
\usepackage{float}
\usepackage{graphicx,wrapfig}
\newcommand*\diff{\mathop{}\!\mathrm{d}}
\newcommand{\vect}[1]{\mathbf{#1}}

\newcommand*\Laplace{\mathop{}\!\mathbin\bigtriangleup}
\ifCLASSINFOpdf
\else
\fi
\begin{document}
%
\title{A General Analytical Approximation to Impulse Response of 3-D Microfluidic Channels in Molecular Communication}
%
%
%

\author{ Fatih Din\c{c},~\IEEEmembership{Student Member,~IEEE},
         Bayram Cevdet Akdeniz,~\IEEEmembership{Student Member,~IEEE}, Ali Emre Pusane,~\IEEEmembership{Member,~IEEE}  and Tuna Tugcu,~\IEEEmembership{Member,~IEEE},  

}

\maketitle

\begin{abstract}
In this paper, the impulse response for a 3-D microfluidic channel in the presence of Poiseuille flow is obtained by solving the diffusion equation in radial coordinates. Using the radial distribution, the axial distribution is then approximated accordingly. Since Poiseuille flow velocity changes with radial position, molecules have different axial properties for different radial distributions. We, therefore, present a piecewise function for the axial distribution of the molecules in the channel considering this radial distribution. Finally, we lay evidence for our theoretical derivations for  impulse response of the microfluidic channel and radial distribution of molecules  through comparing them using various  Monte Carlo simulations. 
\end{abstract}


%
\IEEEpeerreviewmaketitle

\section{Introduction}

Molecular communication via diffusion (MCvD) is one of the most promising areas for nanonetworking due to its 	biocompatability. It is based on encoding information symbols by releasing messenger molecules into a fluidic environment. Released molecules diffuse through the environment under Brownian motion and the receiver makes a decision on the transmitted symbols by \textit{observing} or \textit{absorbing} the released molecules.

There are several different channel models proposed in the molecular communication literature. An extensive survey that involves the compilation of these channel models is presented in \cite{farsad2016comprehensiveSO}. Despite the large number of different diffusion channels in the literature, they all have one common problem due to the nature of the diffusion: inter symbol interference (ISI). Since the movement of molecules are slow and random in Brownian motion, some of the released molecules may not reach to the destination until the desired time. This possibly leads to an adverse effect on decoding. There are many modulation and equalization methods to eliminate the molecules that cause ISI \cite{arjmandi2013diffusion}, \cite{tepekule2015isi}, \cite{kabir2015d}, \cite{arjmandi2017isi}, \cite{akdeniz2018optimal}. In addition to these methods, channel models that diminish  ISI have been proposed by considering the reasons that cause ISI. 

It is clear that, ISI occurs due to the dispersion and slow movement of the molecules. Especially in unbounded environments, the molecules are uniformly dispersed in the space and hence, the number of  molecules reaching  the receiver decreases. Therefore, using barriers in the channel can be a reasonable approach to keep the molecules closer to the destination and is a more realistic channel type considering biomedical applications. As proposed in \cite{gine2009molecular}, vessel-like structures are good candidates for long range molecular communication since they preserve released molecules in a guided range. Another beneficial factor in molecular communication channel for reducing ISI is flow which increases the speed of the molecules \cite{farsad2016comprehensiveSO}. Therefore, using microfluidic channels assisted by flow does not only diminish ISI but also increases the data rate.

\begin{figure*}{h}

\includegraphics[scale=0.47]{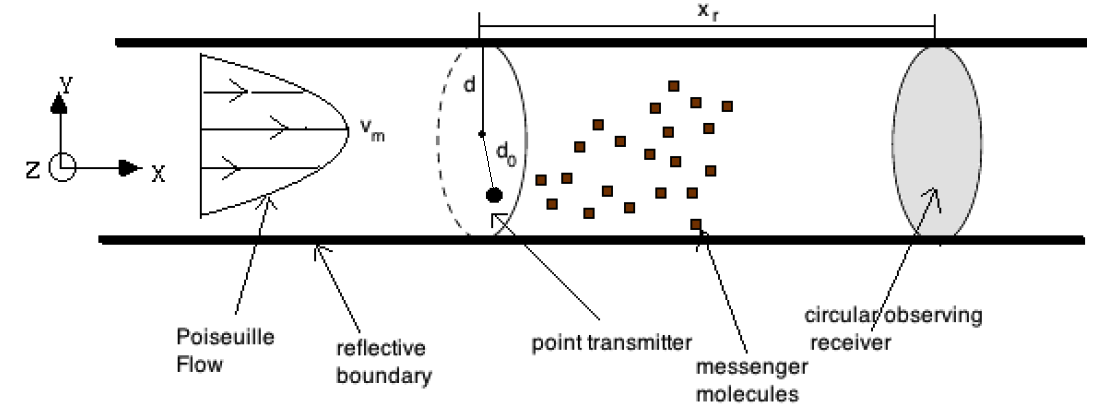} 

\caption{General channel model for a point transmitter and a planar observing receiver. }
\label{fig1}
\end{figure*}

There are various works related to microfluidic  or vessel-like channels. In \cite{felicetti2013establishing}, a decoding scheme for
molecular communications in blood vessels has been proposed. In \cite{bicen2013molecular}, a rectangular microfluidic channel with flow has been modelled and analyzed. In \cite{sun2017channel}, numerical capacity analysis of the vessel like molecular channel with flow is examined. In \cite{turan2018channel}, a partially covering receiver in a vessel-like channel without flow is examined, and the channel characteristics are analyzed. Although the analytical channel impulse responses for unbounded channels are derived, the channel impulse response for microfluidic channel with Poiseuille flow has not been derived yet in the molecular communication literature except for some special cases. In \cite{bicen2013system}, the flow models of microfluidic channels with different cross-section area are presented and the impulse response is derived by solving a 1-D diffusion-advenction equation, which is only valid for some specific cases. In \cite{wicke2017modeling}, channel impulse responses of point and planar transmitter in a microfluidic channel have been derived for dispersion and flow dominated cases. For the first case, radial distribution of the molecules is assumed to be uniform and the system is reduced to 1-D to solve the channel response. For the latter case, only the flow is considered by neglecting the effect of diffusion and the channel impulse response is obtained accordingly. Although that paper includes an elegant and extensive work for microfluidic channels with Poiseuille flow, it is only valid for the uniform radial distribution or flow dominated regions. This assumptions occurs, if Peclet number, a unitless number that compares the effects of flow and diffusion, is much higher or much lower than the ratio of the radius of the channel and the distance between the transmitter and receiver. Therefore, for other cases, derivation of the channel impulse response  still remains as an open problem. In this paper, the analytical channel impulse response of microfluidic channel that involves Poiseuille flow is derived when an arbitrarily placed point transmitter  and a planar observing receiver that fully covers the cross-section of the channel. In order to determine this function,  firstly  the analytical formula of the radial distribution of the molecules in time is derived. Using this distribution,  the average velocity, displacement of a molecule and finally the probability of observation of a molecule by the receiver (which can also be regarded as channel impulse response) are obtained.  All these functions enable us to examine a microfluidic channel without any real time simulation. Furthermore,  using these functions, some channel properties  like required time to reach uniform radial distribution in channel is extracted.

The main contributions of this paper to the literature can be listed as follows:

\begin{itemize}
    \item Derivation of the radial distribution of the released molecules in a vessel-like molecular communication channel under the effects of Poiseuille flow and diffusion.
    \item  A piecewise approximation of the axial distribution of the released molecules for all cases.

\end{itemize}

\section{System Model}

\label{sec2} 

The considered system model is depicted in Fig. \ref{fig1}. For the sake of simplicity, the coordinates in the channel are defined using cylindrical coordinates $(r,x)$, where $r=\sqrt{y^2+z^2}$ $\in$ $\left[0,d \right]$ and $x$ $\in$ $\left[-\infty,\infty \right]$. In this figure, a point transmitter placed at an  arbitrary axial ($x$ direction) and radial ($r$ direction) position denoted by $d_0$, and a  circular planar observing receiver is placed $x_r$ away from the axial position of the point transmitter. The boundaries of the microfluidic channel are reflecting and there is a Poiseuille flow that changes with the axial position as

\begin{equation}
v(r)=v_m \left(1-\frac{r^2}{d^2} \right),
\label{eq:vel}
\end{equation}

\noindent where $v_m$ is the maximum velocity that occurs at the center of the microfluidic channel. The  planar observer receiver observes the number of molecules passing through its surface and makes a decision based on its observations without removing the messenger molecules from the environment. In addition to the flow, diffusion also exists in the channel and  the diffusion process is modelled using the diffusion coefficient $D$, which is related to the variance of the Brownian motion. Since the flow is only available on the $x$ axis, for each ${ \triangle t}$  seconds, the displacement of  a molecule  at a radial position $r$ can be modelled in Cartesian coordinates as

\begin{align}
  &\triangle x={ \triangle t}v(r)+\mathcal{N}\left( 0,2D\triangle t \right),  \nonumber\\
  &\triangle y=\mathcal{N}\left( 0,2D\triangle t \right),  \nonumber\\
  &\triangle z=\mathcal{N}\left( 0,2D\triangle t \right).
  \label{e0}
\end{align}

From Eq. (\ref{e0}), one can easily observe that the distribution of $\triangle y$ and $\triangle z$ (hence, radial displacement $\triangle r$) are not dependent on $\triangle x$, but distribution of $\triangle x$ is dependent on  $\triangle r$. In other words, the movement of the molecules along the radial axis is purely diffusive while movement of the molecules along the axial axis is a combination of the diffusion and flow, whose velocity is determined by radial position. In order to compare the effect of flow and diffusion, Peclet number ($Pe$) is a useful dimensionless metric that can be obtained for the microfluidic channel as

\begin{align}
  Pe=\dfrac{v_m d}{2 D}.
  \label{e11}
\end{align}

\noindent Note that for pure diffusion $Pe=0$, and for pure flow (i.e., advection) $Pe$ approaches 
$\infty$.

\section{Channel Impulse Response}

In order to derive the impulse response of the channel, we need to derive the joint radial, axial, and time distribution of the molecules, $p\left( x,r,t | d_0, d, x_r \right) $, in the microfluidic channel. As can be seen in Eq. (\ref{e0}), the radial distribution is independent of the axial distribution, but the axial distribution is dependent on the radial distribution. Considering this fact, we can rewrite  $p\left( x,r,t | d_0, d, x_r \right)$ as

\begin{align}
 p\left( x,r,t | d_0, d, x_r \right)=p\left(r,t | d_0, d, x_r \right) \cdot p\left( x,t |r, d_0, d, x_r \right).
  \label{e2}
\end{align}

 \noindent Therefore, our aim is  finding the radial distribution $p\left(r,t | d_0, d, x_r \right)$ first, and then using this distribution to obtain the axial distribution $p\left( x,t |r, d_0, d, x_r \right)$. Once $p\left( x,r,t | d_0, d, x_r \right)$ and $p\left(r,t | d_0, d, x_r \right)$ are derived, the channel impulse response of the circular observer is obtained as
 
 \begin{align}
 n_{hit}\left( t | d_0, d, x_r \right)= \frac{\partial}{\partial t}\int_{x_r}^\infty p\left( x,t |r, d_0, d, x_r \right) \diff x.
 \label{nhit}
 \end{align}

 Having found  $n_{hit}\left( t | d_0, d, x_r \right)$, one can easily find the fraction of the observed  molecules by the receiver until time $t$, $N_{hit}\left(t| d_0, d, x_r \right)$, by integrating $n_{hit}\left(t| d_0, d, x_r \right)$ with respect to time as
 
  \begin{align}
 N_{hit}\left( t | d_0, d, x_r \right)= \int _{ 0 }^{ t }{ n_{hit}\left( \tau| d_0, d, x_r \right) } \diff \tau.
 \label{Nhit}
 \end{align}

 \subsection{Derivation of the radial distribution  $p\left(r,t | d_0, d, x_r \right)$  }
 
The radial distribution can be obtained by simply solving the diffusion equation or by drawing comparison to the heat flow, as discussed in \cite{carslaw1959conduction}. Here, we shall find the radial distribution of molecules using the former method. To describe the diffusion of the molecule inside the radial region, a solution to the Fick's Law, satisfying the necessary boundary conditions is needed. The equation is given as 
\begin{equation} \label{eq:fick}
D \grad^2 P\left(r,t | d_0, d, x_r \right) = \frac{\partial P\left(r,t | d_0, d, x_r \right)}{\partial t},
\end{equation}
where $P\left(r,t | d_0, d, x_r \right)$ is the probability density of the molecule. The boundaries are reflecting, meaning that the probability current normal to the boundaries should be zero. Furthermore, the molecule is assumed to be situated at a distance $d_0$ away from the origin for $t=0$, which results in the following two boundary conditions:
\begin{subequations} \label{eq:boundary}
\begin{align}
\frac{\partial P\left(r,t | d_0, d, x_r \right)}{\partial r}\Big|_{r=d} &= 0, \label{eq:neuman} \\
P\left(r,t | d_0, d, x_r \right)\Big|_{t=0}&=\frac{\delta(r-d_0)}{2 \pi r} \label{eq:dirac},
\end{align}
\end{subequations}
where we recall that, under Neumann boundary conditions, the Laplacian operator ($\Laplace$) is guaranteed to have a unique solution up to an addition of a constant, which can be regarded as the normalization constant. In order to solve $P\left(r,t | d_0, d, x_r \right)$ seperation of variable anatsz is used as

\begin{equation}
P\left(r,t | d_0, d, x_r \right)=\phi(r,\theta) T(t),
\end{equation}
which leads to the equation
\begin{equation*}
D \frac{\Laplace \phi(r,\theta)}{\phi(r,\theta)} = \frac{T'(t)}{T(t)}= - \mu^2, 
\end{equation*}
from which one can easily deduce that:
\begin{equation}
T(t) = C e^{-\mu^2 t}, \label{eq:time}
\end{equation}
and arrive at the Neumann-eigenvalue problem for the Laplacian operator:
\begin{equation}
 \Laplace \phi(r,\theta) = - \frac{\mu^2}{D} \phi(r,\theta).
\end{equation}

At this point, it is important to recall some key properties of Laplacian  operator \cite{grebenkov2013geometrical,szego,weinberger}. The eigenvalues $ \mu^2/D$ are non-negative and real, as well as the eigenvectors corresponding to distinct eigenvalues being orthogonal and forming a basis for all possible solutions. The non-negativity of the eigenvalues ensures that $T(t)$ does not tend to infinity as $t \to \infty$, whereas the orthogonal basis guarantees a unique solution. Here, we invoke the idea of angular symmetry (SO(2) symmetry) in the system. Due to SO(2) symmetry, the position-dependent part of the ansatz depends only on the distance from the origin and not the angle-$\theta$, i.e.,  $\phi(r,\theta)=\phi(r)$. This choice eliminates certain eigenvalues and corresponding eigenvectors, from the solution. Nonetheless, the coefficients corresponding to the non-symmetrical eigenvectors are zero due to the symmetry of the system,  removing our burden for further calculations.

Rewriting the eigenvalue equation in polar coordinates, we obtain:
\begin{equation*}
r^2\phi''(r)+ r\phi'(r) + \frac{\mu^2}{D} r^2 \phi(r)=0,
\end{equation*}
where $\phi'(r)$ denotes the derivative of $\phi(r)$ with respect to $r$. The most general solution is
\begin{equation*}
\phi(r) = J_0\left(\frac{\mu}{\sqrt{D}} r\right) + c Y_0\left(\frac{\mu}{\sqrt{D}} r\right),
\end{equation*}
where $J_n$ and $Y_n$ are the Bessel's function of the first and second kind, respectively and $c$ is a constant to be determined by the boundary conditions. Here, the coefficient of $J_0$ is chosen arbitrarily as the overall coefficient $C$ of the solution is lumped into the time dependent part in (\ref{eq:time}). The  solution can now be shaped according to the boundary conditions given in Eq. (\ref{eq:boundary}).

One important observation is that, for $t>0$, the probability density function $P\left(r,t | d_0, d, x_r \right)$ does not diverge for $\vect r= (0,0)$, resulting in $c_{2}=0$. The most general solution is then of the form: 
\begin{equation*}
P\left(r,t | d_0, d, x_r \right)= \sum_{n=0}^\infty C_{n} J_0\left(\frac{\mu_n}{\sqrt{D}} r\right) e^{-\mu_n^2 t},
\end{equation*}
where $C_n$ and $\mu_n$ are to be specified by the boundary conditions. We arbitrarily define the starting index as $n=0$. Once the boundary condition given in Eq. (\ref{eq:neuman}) is invoked, we arrive at
\begin{equation*}
\frac{\partial P\left(r,t | d_0, d, x_r \right)}{\partial r} \Big|_{r=d}=0 \implies J_1\left(\frac{\mu_n}{\sqrt{D}} d \right) = 0.
\end{equation*}

The first order Bessel function $J_1(r)$ has infinitely many zeros. These zeros, defined as $\beta_n= \frac{\mu_n}{\sqrt{D}} d$, then constitute infinitely many terms for the solution. With the notation $\beta_0=0$, the solution is a sum of infinitely many terms given as

\begin{equation*}
P\left(r,t | d_0, d, x_r \right)= \sum_{n=0}^\infty C_{n} J_0\left(\beta_n \frac{r}{d}\right) e^{-\mu_n^2 t}.
\end{equation*}

Before imposing the final condition given in Eq. (\ref{eq:dirac}), it is useful to note the following normalization identity for Bessel functions \cite{PDE}, which is given as
\begin{equation*}
\int_0^1 x J_0(\beta_n x) J_0(\beta_m x) \diff x = 0.5 J_0(\beta_n)^2 \delta_{nm},
\end{equation*}
where $\delta_{nm}$ is the Kronecker delta function.
Using this identity for the initial condition given in Eq. (\ref{eq:dirac}), we conclude that
\begin{equation*}
C_n = \frac{J_0\left(\beta_n \frac{d_0}{d}\right)}{\pi d^2 J_0^2(\beta_n) },
\end{equation*}
from which we find the final solution to be
\begin{equation}
P\left(r,t | d_0, d, x_r \right)= \sum_{n=0}^\infty \frac{J_0\left(\beta_n\frac{d_0}{d}\right)}{\pi d^2J_0^2\left(\beta_n \right)} J_0\left( \frac{\beta_n}{d} r\right) e^{- \beta_n^2\frac{D t}{d^2} },
\label{radial}
\end{equation}
 where the first term ($\beta_0=0$) corresponds to the uniform distribution $P\left(r,t | d_0, d, x_r \right) \to 1/\pi d^2$, as $t\to \infty$. Practically, it takes much shorter time to reach the uniform distribution. Considering $P\left(r,t | d_0, d, x_r \right)$ in Eq. (\ref{radial}), while $n=0$ corresponds to the uniform distribution, $n=1$ is the dominating term that makes  $P\left(r,t | d_0, d, x_r \right)$ non-uniform. Therefore, if this term is arbitrarily small that can be assumed to be effectively zero, we can find the required time that radial distribution becomes uniform as
 
  \begin{equation}
 ke^{ -\beta _{ 1 }^{ 2 }\frac { Dt }{ d^{ 2 } }  }<\epsilon,
 \label{bound}
  \end{equation}
 where we define the bound $\left| \frac { J_{ 0 }\left( \beta _{ 1 }\frac { d_{ 0 } }{ d }  \right)  }{ J_{ 0 }^{ 2 }\left( \beta _{ 1 } \right)  } J_0\left( \frac{\beta_1}{d} r\right) \right| \leq k$ and $\epsilon/\pi d^2$ as the allowed deviation in the radial distribution from the uniformity. Here, we note that $\pi d^2$ is not lumped into the parameter $\epsilon$ to keep the measures unitless, since as $t \to \infty$, $P\left(r,t | d_0, d, x_r \right) \to 1/\pi d^2$ and $\epsilon$ can be interpreted as the fractional error. Therefore, $P\left(r,t | d_0, d, x_r \right)$ becomes uniform if the following condition is satisfied:
   \begin{equation}
t \geq \frac { d^{ 2 } }{ D \beta_1^2} \log( { \cfrac { k }{ \epsilon  }  }).
 \label{bound2}
  \end{equation}
We note that this bound is in line with the findings of \cite{taylor1953dispersion}. Taking  $\epsilon=10^{-2}$, we obtain
 
    \begin{equation}
    t \geq\frac { d^{ 2 } }{ D \beta_1^2 }(log(100)+log(k)).
 \label{bound3a}
  \end{equation}
  
  Since $\beta_1 \approx 3.83$ and $k$ is on the order of 1, a bound on time for uniform radial distribution can be approximated as
  
      \begin{equation}
  t \geq t^{*} \approx\frac { d^{ 2 } }{ 3 D }.  
 \label{bound3}
  \end{equation}

  Therefore, for $t \geq t^{*}$, we can expect radial 	homogeneity. Having found the probability density function $P\left(r,t | d_0, d, x_r \right)$, we can find the radial distribution as
 \begin{equation}
     p\left(r,t | d_0, d, x_r \right)=2\pi rP\left(r,t | d_0, d, x_r \right).
 \end{equation}

    \begin{figure*}[h]
\centering
\subfigure[$d=5\mu m$, $d_0=0$, $P_e=250$, $P_c=4000$ ]{
\includegraphics[width=.31\textwidth]{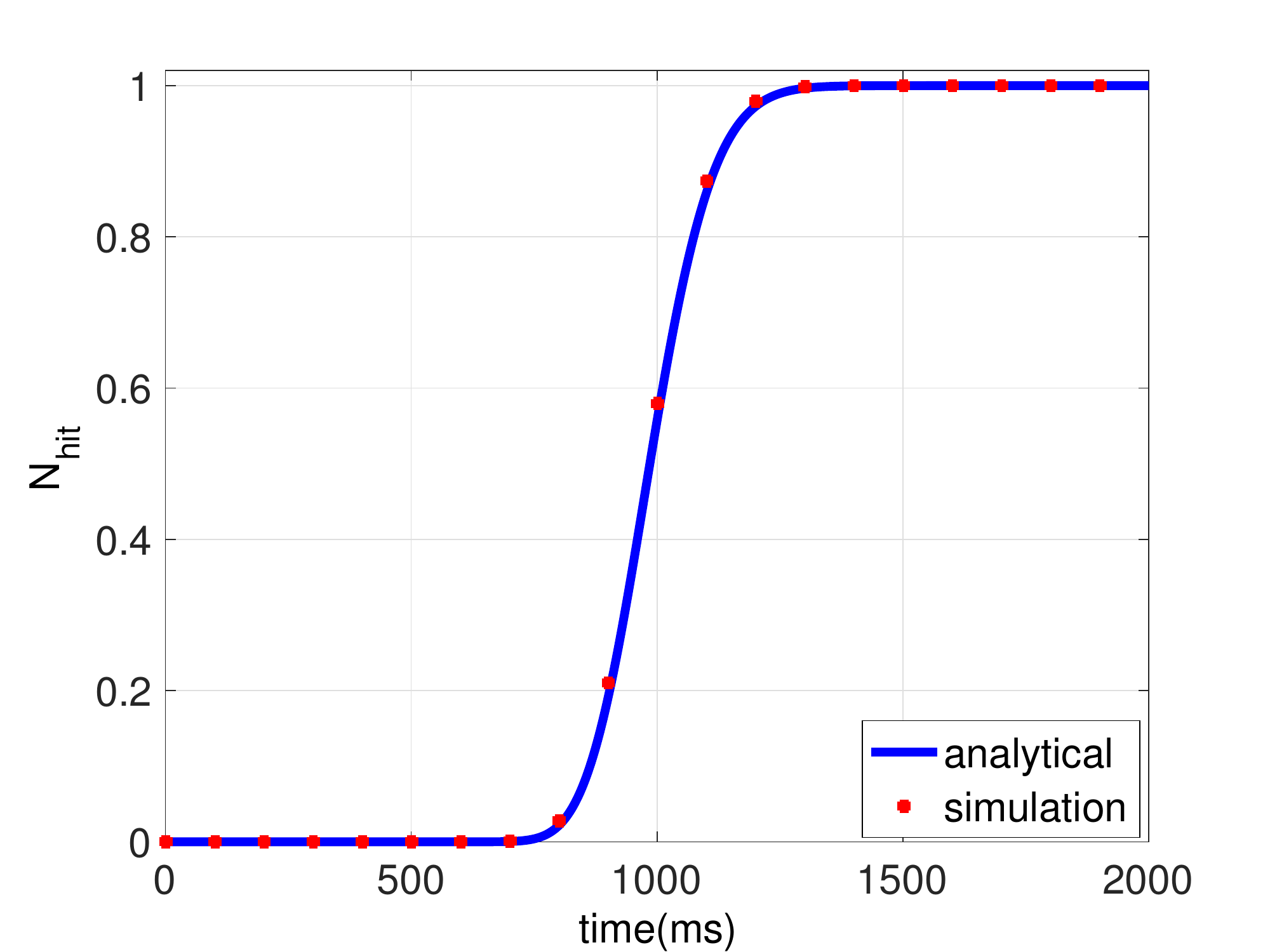}
\label{fig:rad1}
}
\subfigure[$d=15\mu m$, $d_0=0$, $P_e=750$, $P_c=1333$, $t^*=3/4 s$  ]{
\includegraphics[width=.31\textwidth]{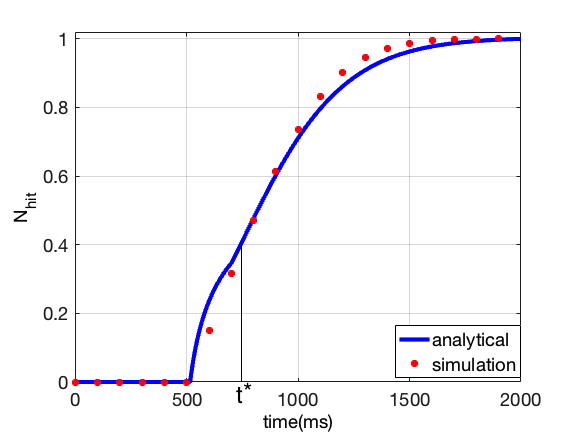}
\label{fig:rad2}
}
\subfigure[$d=20\mu m$, $d_0=0$, $P_e=1000$, $P_c=1000$, $t^*=4/3 s$ ]{
\includegraphics[width=.31\textwidth]{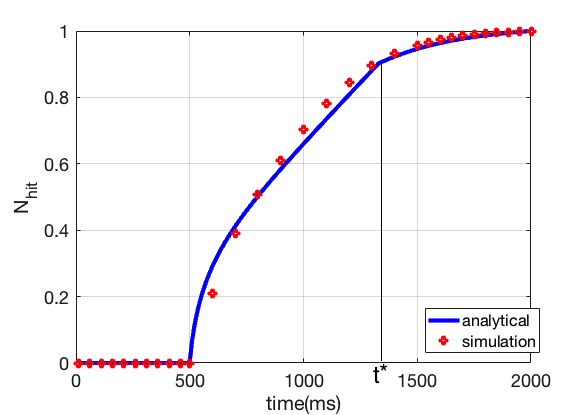}
\label{fig:rad3}
}
\subfigure[$d=100 \mu m$, $d_0=0$, $P_e=5000$, $P_c=200$]{
\includegraphics[width=.31\textwidth]{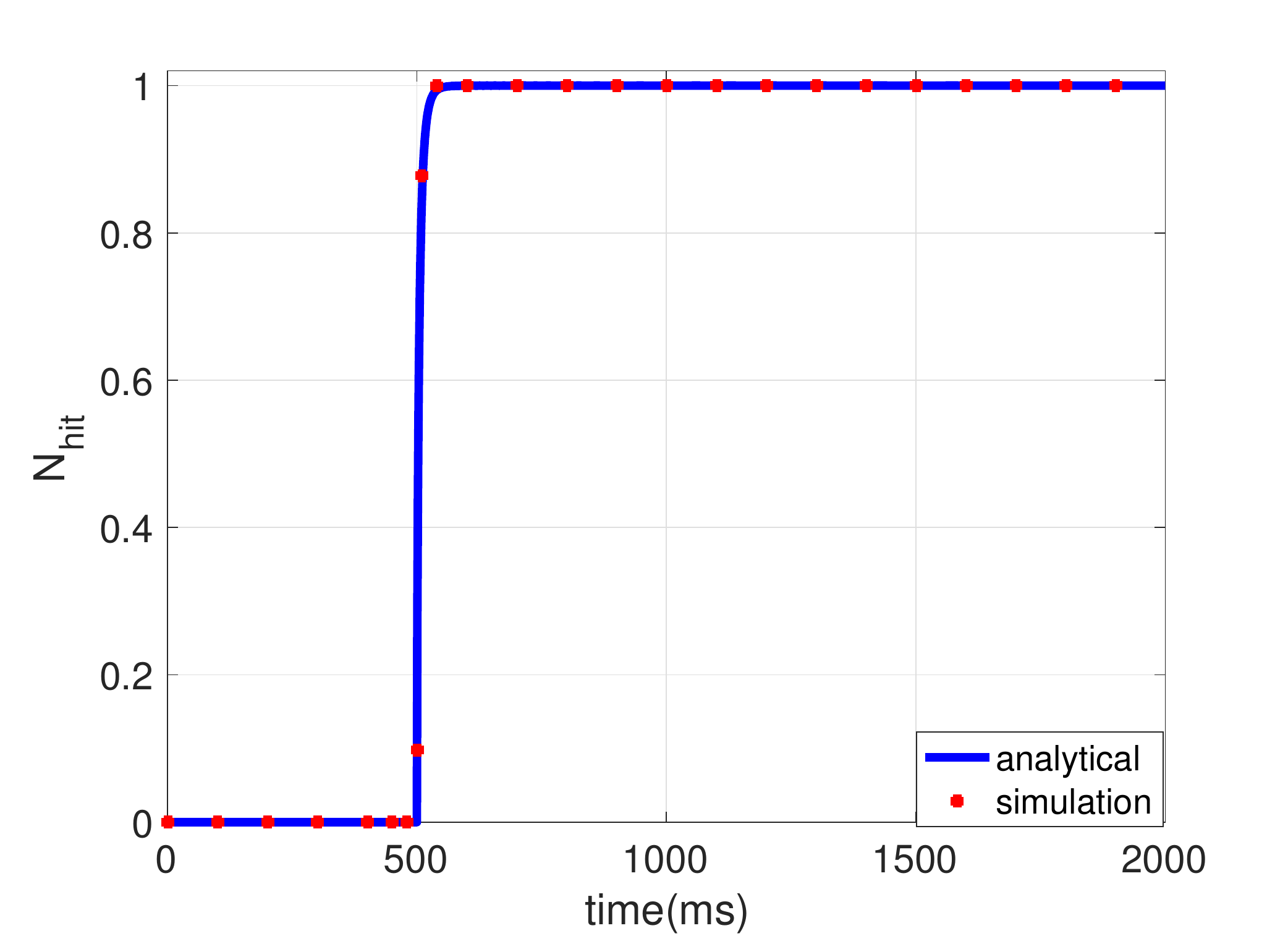}
\label{fig:rad4}
}
\subfigure[$d=15 \mu m$, $d_0=0.5d$ $P_e=750$, $P_c=1333$ ]{
\includegraphics[width=.31\textwidth]{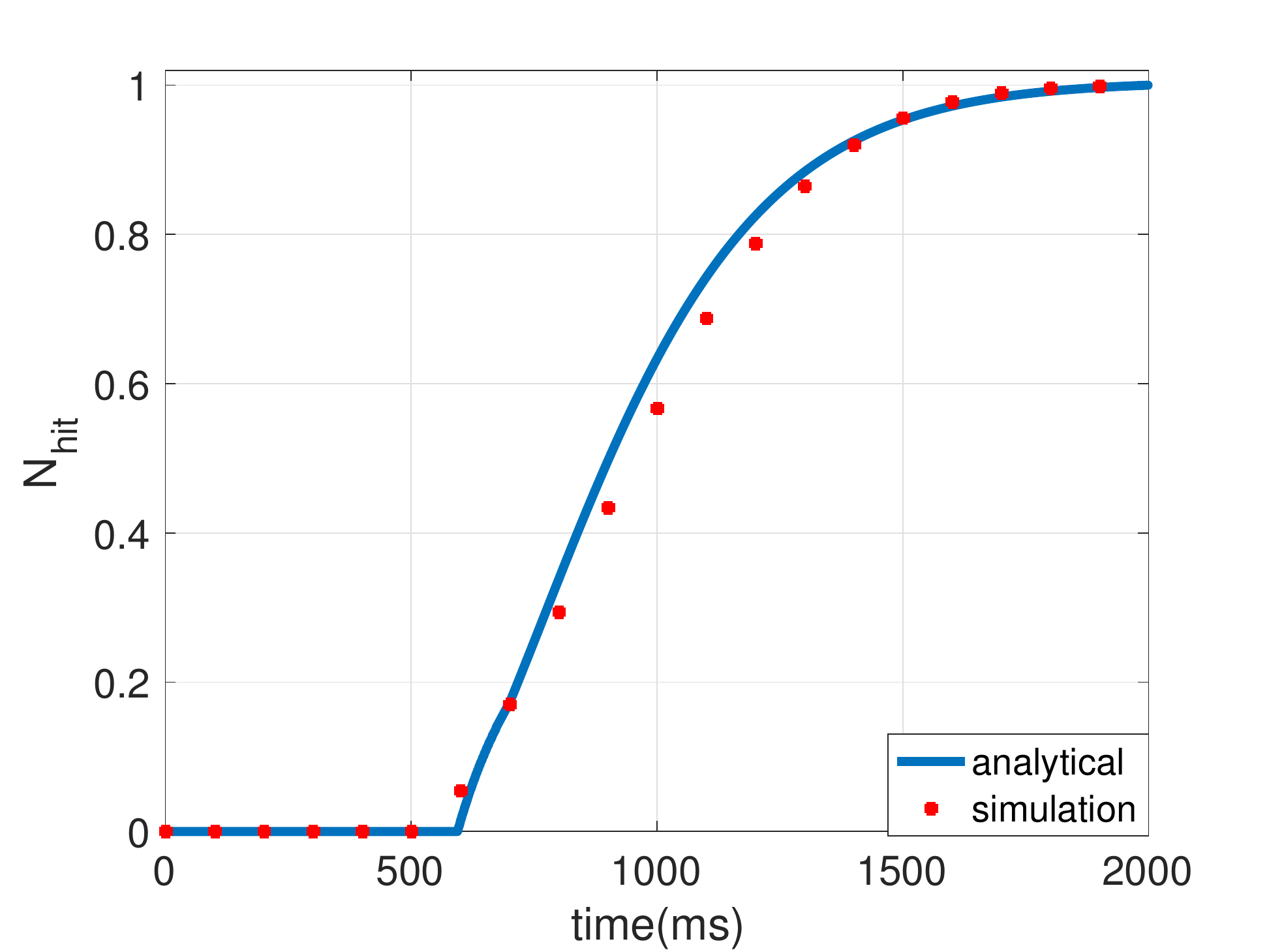}
\label{fig:rad5}
}
\subfigure[$d=40 \mu m$, $d_0=0.25d$ $P_e=2000$, $P_c=500$ ]{
\includegraphics[width=.31\textwidth]{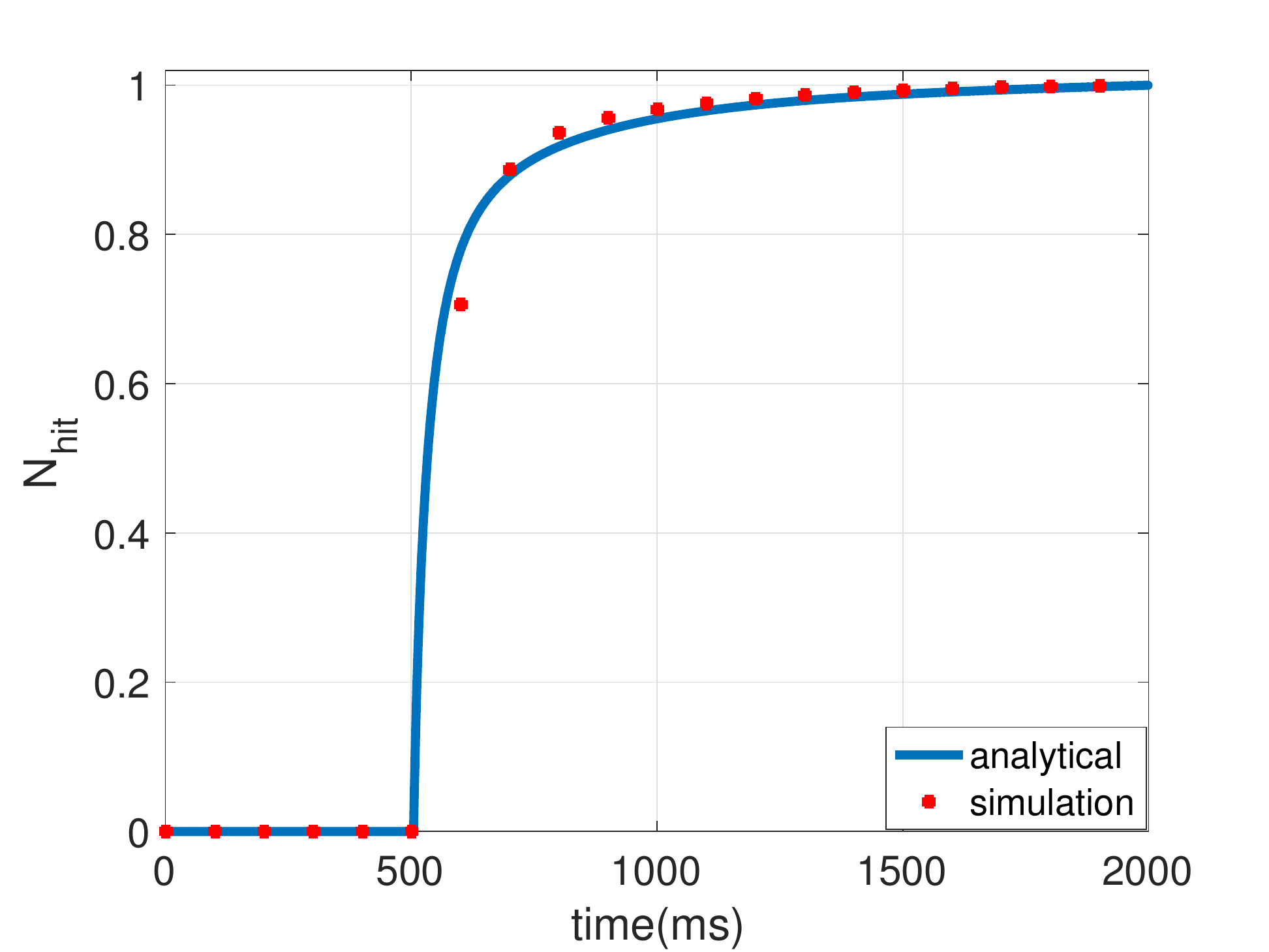}
\label{fig:rad6}}

\caption{Simulation and analytical cumulative axial molecule distribution, $N_{hit}\left( t | d_0, d, x_r \right)$ for   $D=100 \times 10^{-12} \mu m / s^2 $, $x_r=5 mm$, $v_m=10mm/s $ and different $d$ values which indicate different $P_e$ regions. }
\label{fig:rad}
\end{figure*}

\begin{figure*}[h]
\centering

\subfigure[$d=15\times 10^{-6}m$, $t=200 ms$]{
\includegraphics[width=.25\textwidth]{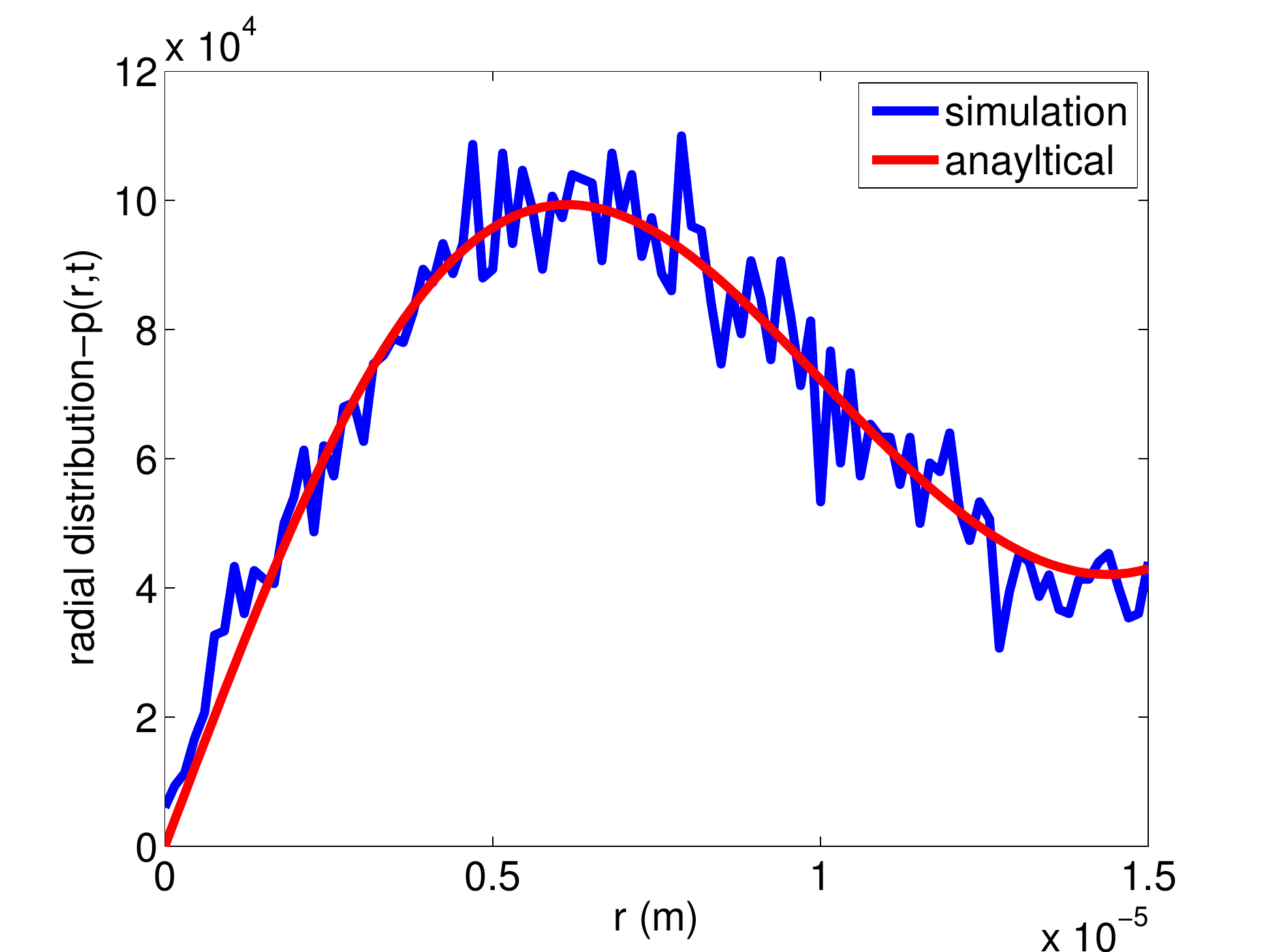}
\label{fig:BERa2}
}
\hspace{-0.9cm}
\subfigure[$d=15\times 10^{-6}m$, $t=2s$]{
\includegraphics[width=.25\textwidth]{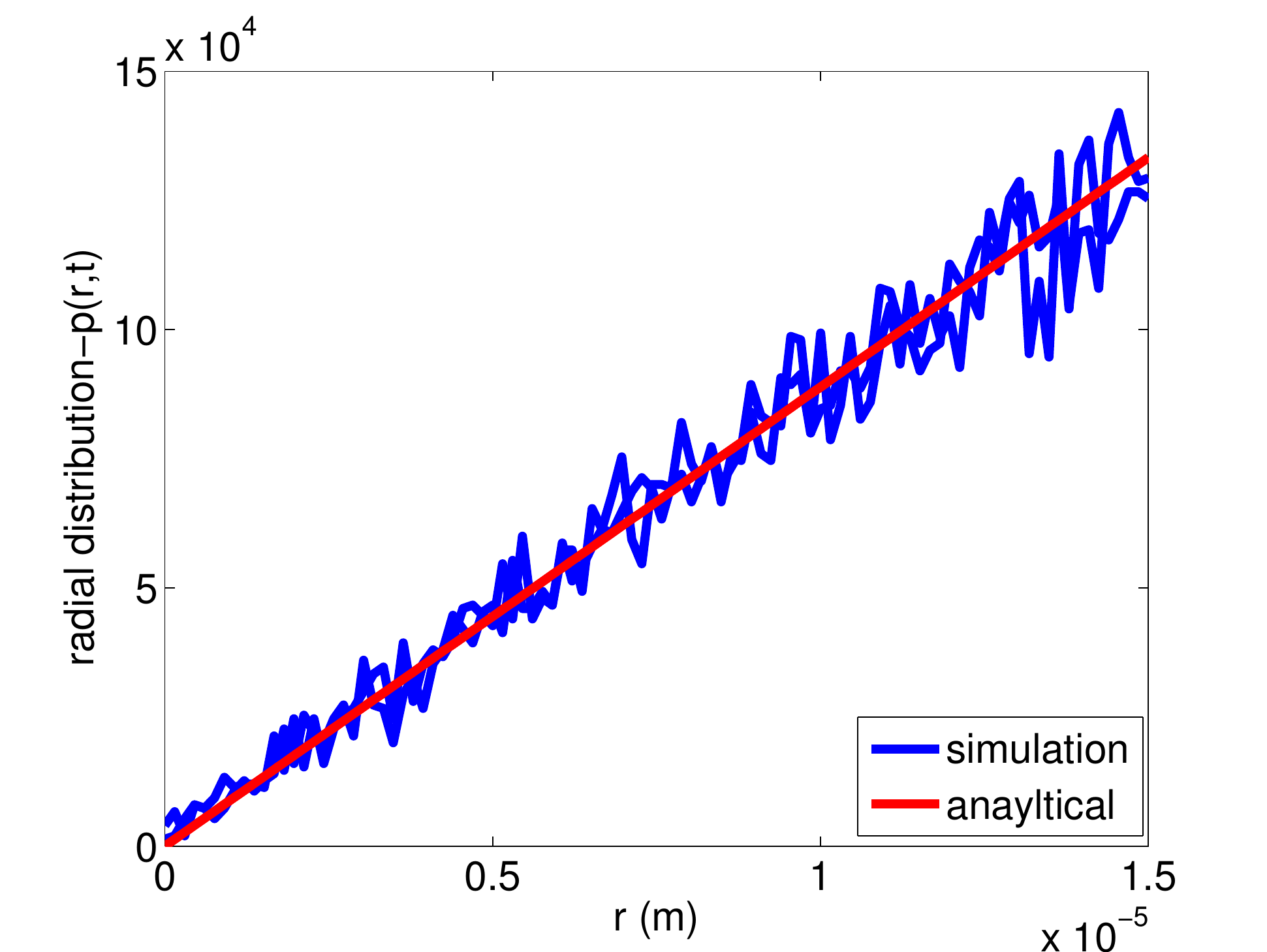}
\label{fig:BERb2}
}
\hspace{-0.9cm}
\subfigure[$d=40\times 10^{-6}m$, $t=2s$ ]{
\includegraphics[width=.24\textwidth]{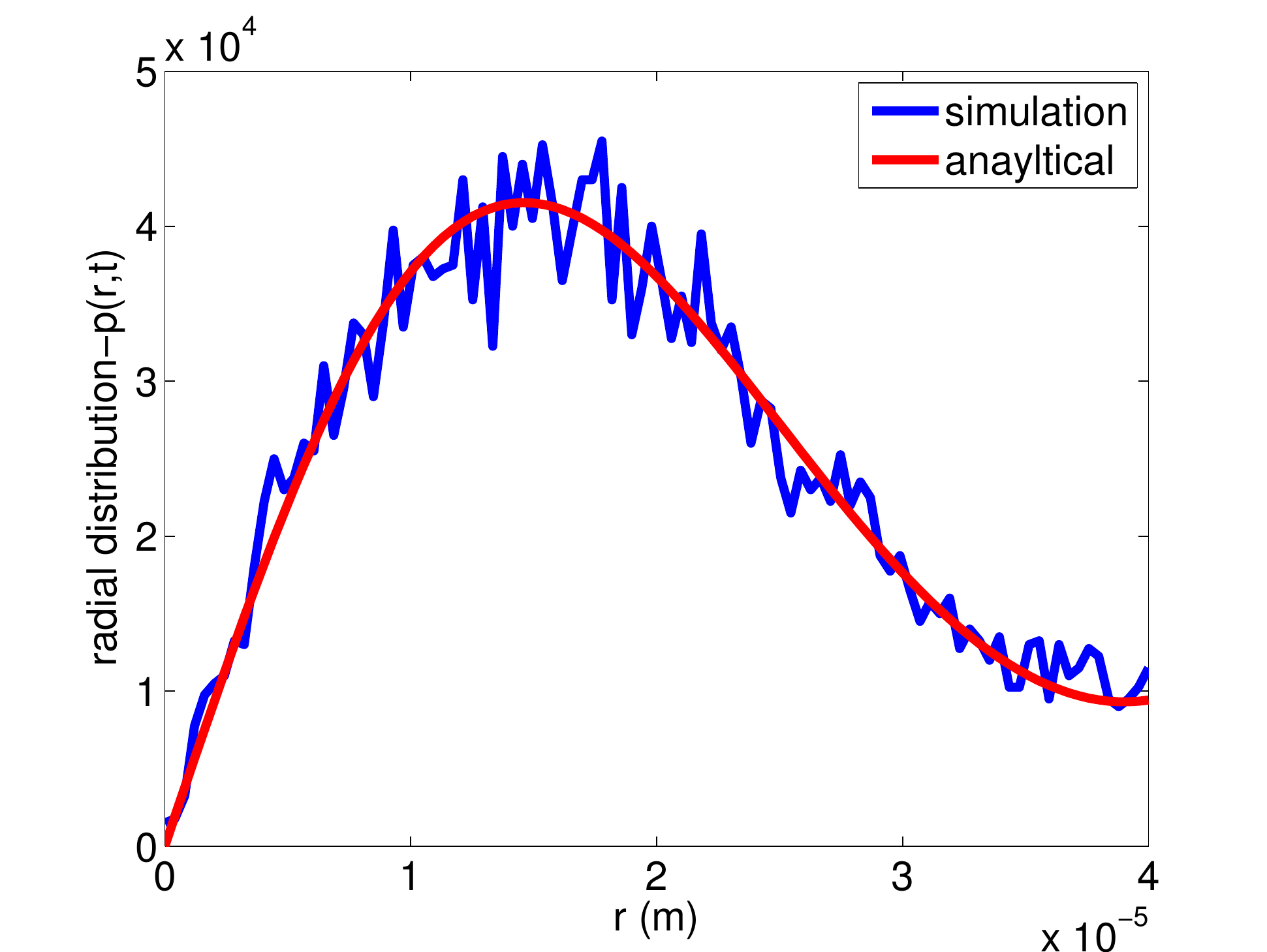}
\label{fig:BERc2}
}
\hspace{-0.9cm}
\subfigure[$d=40\times 10^{-6}m$, $t=8s$]{
\includegraphics[width=.24\textwidth]{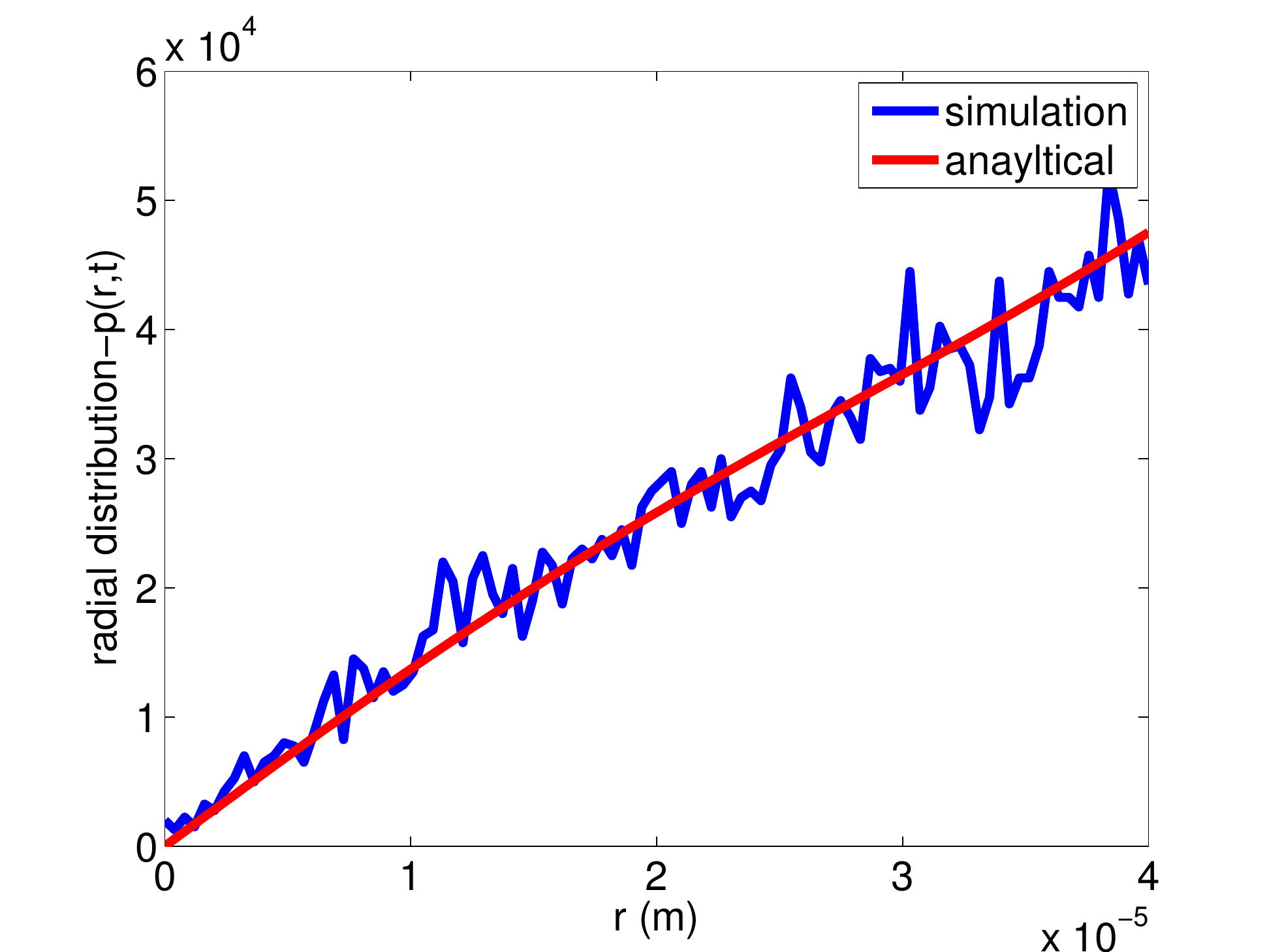}
\label{fig:BERd2}
}

\caption{Simulation and analytical  radial molecule distribution, $p\left( r,t | d_0, d, x_r \right)$ for   $D=100\times 10^{-12}  m / s^2 $, $x_r=5 mm$, $v_m=10mm/s $ and different $d$ values and time $t$. Noting that, $p\left(r,t | d_0, d, x_r \right)=2\pi rP\left(r,t | d_0, d, x_r \right)$, linearity on $p\left( r,t | d_0, d, x_r \right)$ implies uniformity on $P\left( r,t | d_0, d, x_r \right)$. }
\label{fig:BER2}
\end{figure*}

 \subsection{Derivation of the axial distribution $p\left( x,t |r, d_0, d, x_r \right)$}
 
 The axial distribution of the molecules in the microfluidic channel has different characteristics for different $p\left(r,t | d_0, d, x_r \right)$. In particular, as discussed in \cite{probstein2005physicochemical}, if $p\left(r,t | d_0, d, x_r \right)$ is uniformly distributed, $p\left( x,t |r, d_0, d, x_r \right)$ can be easily identified as

  \begin{equation}
     p\left( x,t |r, d_0, d, x_r \right)=\cfrac { 1 }{ \sqrt { 4\pi {D}_{e}t }  } \exp( -\cfrac { \left( x-\cfrac { { v }_{ m }t }{ 2 }  \right)^2  }{4{D}_{e}t}  ), 
     \label{axial}
 \end{equation}

 \noindent which is equivalent to  $\mathcal{N}
 \left( \cfrac { { t v }_{ m } }{ 2 } ,2{D}_{e}t \right)$ and can be considered as a 1-D Brownian motion with effective diffusion coefficient ${D}_{e}=D\left( 1+\cfrac { { P }_{ e }^{ 2 } }{ 48 }  \right)$ shifted with  average axial displacement for any time $t₺$.
 
 The fact that Eq. (\ref{axial}) is valid  for uniformly distributed $p\left(r,t | d_0, d, x_r \right)$ implies it is valid when following condition is satisfied,

  \begin{equation}
{ P }_{ e }\ll { P }_{ c }=\cfrac { 4 x_r }{ d }, 
   \end{equation}
    
  \noindent as explained in  \cite{probstein2005physicochemical}. This implication makes sense since in order to obtain a uniform distribution in radial space, the channel should have either small radius ($d$) or high diffusion coefficient ($D$) so that the molecules disperse in the channel rapidly to reach the uniform state. 
  
  On the other hand, as ${ P }_{e}$ increases, it takes some time for $p\left(r,t | d_0, d, x_r \right)$  to become uniform. Hence, during this period, Eq.(\ref{axial}) cannot be used. We therefore propose two different regions, namely uniform and non-uniform radial regions, and using these regions a partial function for the derivation of $p\left( x_r,t |r, d_0, d, x_r \right)$ is obtained:
    
    \subsubsection{$p\left( x_r,t |r, d_0, d, x_r \right)$ for uniform radial distribution range}
    
    As stated before, $p\left( x,t |r, d_0, d, x_r \right)$ can be written using Eq. (\ref{axial}), if radial distribution is uniform. We have already shown that for $t>t^*$ the radial distribution is uniform. Therefore for this period we can present the axial distribution using Eq. (\ref{axial}). On the other hand since the average displacement is different for $t>t^*$ due to non-uniform radial distribution, we should evaluate the average displacement and plug it to Eq. (\ref{axial}) instead of $\cfrac { { v }_{ m }t }{ 2 }$. For this aim, the average axial displacement of a molecule $ { x }_{ exp }(t)$ for  $t$ seconds can be evaluated using  $p\left(r,t | d_0, d, x_r \right)$ as
 
 \begin{align}
 { x }_{ exp }(t)=\int _{ 0 }^{ t }{ \int _{ 0 }^{ d }{ p\left(r,\tau | d_0, d, x_r \right) v(r) \diff r \diff \tau }  }. 
  \label{xExp}
\end{align}

 Once ${ x }_{ exp }(t)$ is obtained, $p\left( x_r,t |r, d_0, d, x_r \right)$ is obtained using Eq.(\ref{axial}) as

   \begin{align}
     p\left( x_r,t |r, d_0, d, x_r \right)=\cfrac { \exp\left( -\cfrac { \left( x_r- { x }_{ exp }(t)  \right)^2  }{ 4{D}_{e}t }  \right) }{ \sqrt { 4\pi {D}_{e}t }  } ={ f }_{ X_t }({ x }_{ r }). 
     \label{axial2}
 \end{align}
 where ${ f }_{ X_t }({ x }) \sim \mathcal{N}({ x }_{ exp }(t), 2{D}_{e}t )$.

    \subsubsection{$p\left( x_r,t |r, d_0, d, x_r \right)$ for non-uniform radial distribution range}
 
 When ${ P }_{ e }$ is comparable to or greater than ${ P }_{ c }$, the released molecules need some time to reach a uniform radial distribution. Until that time, dispersion of molecules is limited; hence, a new axial distribution model should be defined. Let $v(t)$ be the average velocity and $v_{a}(t)$ be the required average velocity to traverse a distance $x_r$ at time $t$, respectively. Then, one can easily define the following relation:
 
 \begin{equation}
    v_{a}(t)=\frac{x_r}{t}.  
\end{equation}
   
   \noindent Therefore, any molecule whose average velocity is higher than $v_{a}(t)$ until time $t$  passes through the receiver. Considering this fact, the probability of exceeding $x_r$ distance until time $t$ can be written as
   
\begin{equation}
Prob\left( x(t)\ge { x }_{ r } \right) =Prob\left( v(t)\ge v_{ a }(t) \right).      
\end{equation}
where $x(t)$ is a random variable that defines the axial distribution of a molecule and this distribution is not known. Alternatively, one can calculate the probability of exceeding the average required velocity $v_{ a }(t)$ for a given time $t$.  Note that, the required average velocity involves two terms coming from the flow and diffusion as
   \begin{equation}
v_{ a }(t)=v_m \left(1-\frac{{ r }^{ * }(t)^{2}}{d^2}\right)+\frac{U}{t},
\end{equation}
where ${ r }^{ * }(t)$ is the required radial position to achieve $v_{ a }(t)$ with $U$, which is the axial displacement component coming from the diffusion distributed with $f_U(u) \sim \mathcal{N}(0,2Dt)$.
   
  Using $p\left(r,t | d_0, d, x_r \right)$, we can obtain $Prob\left( v(t)\ge v_{ a }(t) \right)$ considering the velocity terms coming from the flow and diffusion using  $p\left(r,t | d_0, d, x_r \right)$ as  
  
    \begin{align}
& Prob\left( v(t)\ge v_{ a }(t) \right) = \nonumber\\
&\int\displaylimits_{-\infty}^{\infty}   \int\displaylimits_{0}^{{ r }^{ * }(t)} \frac{1}{t} \int _{ 0 }^{t} { p\left(r,\tau | d_0, d, x_r \right) d\tau dr f_U(u) du}, 
\label{cuma}
 \end{align}
  \noindent where the result of the first two integral in (\ref{cuma}) gives the average velocity distribution of a molecule due to flow and the outermost integral evaluates the contribution of the diffusion.
  
   Once $Prob\left( x(t)\ge { x }_{ r }\right)$ is obtained, the axial distribution $p\left( x,t |r, d_0, d, x_r \right)$ can also be obtained using this probability as
   \begin{equation}
p\left( x,t |r, d_0, d, x_r \right) =  \frac{\partial}{\partial x_k} \left( 1-Prob\left( x(t)\ge { x }_{ k }\right) \right) \Big|_{x_k=x},    
\end{equation}
where we find the probability density function $p\left( x,t |r, d_0, d, x_r \right)$ from the corresponding cumulative distribution function $Prob\left( x(t)\le { x }\right)$. 
   
 Finally, using (5),  $ n_{hit}\left(t | d_0, d, x_r \right)$  can be partially represented as
    \begin{equation}
  n_{hit}\left(t | d_0, d, x_r \right)=  \begin{cases} 
      \cfrac { \partial  }{ \partial t } Prob\left( v(t)\ge v_{ a }(t) \right), & t\leq t^* \\
      \frac{\partial }{ \partial t}\int_{x_r}^\infty { f }_{ X_t }({ x }) \diff x &t> t^*
   \end{cases}
    \end{equation}

    Accordingly,  $N_{hit}\left( t | d_0, d, x_r \right)$ can be obtained as 
      
       \begin{equation}
  N_{hit}\left( t | d_0, d, x_r \right)=  \begin{cases} 
        Prob\left( v(t)\ge v_{ a }(t) \right) & t\leq t^* \\
     1- { F }_{ X_t }({ x }_{ r }) & t> t^*
   \end{cases}
   \label{finalP}
    \end{equation}
     
\noindent where ${ F }_{ X_t }({x}_{r})$ is the cumulative normal distribution whose mean and variance are ${ x }_{ exp }(t)$ and  $2{D}_{e}t$, respectively. Even though the derivations from (25)-(28) are mainly heuristic and cumbersome, $N_{hit}\left( t | d_0, d, x_r \right)$ is easy to evaluate and to compare with the simulation results, hence is of significant interest for the verification of our findings.

  


\section{Simulation Results}

We have verified the derived cumulative distribution of molecules observed by the receiver, $N_{hit}\left( t | d_0, d, x_r \right)$ and the radial distribution of molecules at the channel $p\left(r,t | d_0, d, x_r \right)$ using Monte Carlo simulations. For both functions, the comparisons are made for three cases:

\begin{enumerate}
    \item ${ P }_{ e }\ll { P }_{ c }$
    \item ${ P }_{ c }\ll { P }_{ e }$
    \item ${ P }_{ e }\sim { P }_{ c }$,
\end{enumerate}

 \noindent namely  ${ P }_{ e }$, is much greater than ${ P }_{ c }$, ${ P }_{ c }$ is much greater than ${ P }_{ e }$ and ${ P }_{ e }$ is comparable with ${ P }_{ c }$, respectively. In particular simulation parameters are listed in Table \ref{my-label}. For all simulations, $10^{5}$ molecules are released and their positions are updated  every $\triangle t=10^{-3}s$ using Eq. (\ref{e0}).

\begin{table}[h]
\centering
\caption{Parameters}
\label{my-label}
\begin{tabular}{|l|l|l|l|l|l|l|}
\hline
$\triangle t =10^{-3}s$ & $d$ & $x_r$ & $v_m$ &$D$ & $P_e$ & $P_c$ \\ \hline
 i) ${ P }_{ e }\ll { P }_{ c }$& $5$ $\mu m$  &   $ 5mm $   &  $10mm/s$ &$10^{-10} m / s^2 $    &$ 250$     & $4000$    \\ \hline
 iii) ${ P }_{ c }\sim { P }_{ e }$& $ 15 \mu m $& $5mm$    & $ 10mm/s $ &$  10^{-10} m / s^2 $   & $ 750 $  & $1333$    \\ \hline
 iii) ${ P }_{ e }\sim { P }_{ c } $& $ 40 \mu m$  & $ 5mm $     & $10mm/s$ &$10^{-10} m / s^2 $    & $2000 $     & $500 $    \\ \hline
  ii)$ { P }_{ c }\ll { P }_{ e } $& $ 100 \mu m$  & $ 5mm$      & $10mm/s$ &$ 10^{-10} m / s^2 $   & $5000$    & $200 $     \\ \hline 
\end{tabular}
\end{table}

In Fig. \ref{fig:rad}, the derived analytical expression $N_{hit}\left( x_r,t | d_0, d, x_r \right)$ is verified using Monte Carlo simulations. As can be seen from the figure, the derived formula fits with simulations for all three regions. Especially for Fig. \ref{fig:rad2} and \ref{fig:rad3}, the proposed piecewise function, which is separated by $t^*$, can be easily identified. For the time duration before $t^*$, the first equation in (\ref{finalP}) is used while the second equation is used for the time duration after $t*$. On the other hand for other simulations $t^*$ is either too small or too high so that only one function is used.  Furthermore, as $d$ (hence $P_e$) increases, the time needed for the molecules to pass through the receiver decreases and converges to $x_r/v_m$. This is expected, since as the radius of the channel $d$ increases, the radial position of the molecules does not disperse so much in the beginning. Hence, the flow speed affecting these molecules is around $v_m$. On the other hand, as $d$ decreases, the molecules can reach the boundary of the channel rapidly, which reduces the velocity of the flow, and thus, it takes more time to reach the receiver. Furthermore, as indicated in Fig. \ref{fig:rad5} and \ref{fig:rad6}, the derived expression is verified for different initial radial positions $d_0$. Another interesting observation from these figures can be obtained by comparing them with  Fig. \ref{fig:rad2} and \ref{fig:rad3}, respectively. As the initial radial position moves from center to boundary, in other words as $d_0$ increases, the initial speed of the molecules decreases. Hence, it takes more time to reach the receiver compared to the $d_0=0$ case. Nonetheless, for all cases, all molecules will be observed by the receiver in a short duration compared to the other channel models in the literature due to flow.

In Fig. \ref{fig:BER2}, the radial distributions for different channel parameters and time are presented both theoretically and numerically. As can be seen from these figures, our proposed analytical formula is verified with the simulations. Furthermore, as indicated in Figs. \ref{fig:BERb2} and \ref{fig:BERd2}, for higher times, $p\left(r,t | d_0, d, x_r \right)$ turns out to be linear; hence, there is a uniform radial distribution. In particular, for the parameters in Fig. \ref{fig:BERa2} and \ref{fig:BERb2}, at $t=200ms$, the radial distribution is non-uniform while at t=$2s$, this distribution becomes uniform. This is expected since for these parameters, $t^{*}=\frac { d^{ 2 } }{ 3D } \approx 0.75s$, resulting a non-uniform distribution for the time below $t^{*}$. The similar situation can also be observed for Fig. \ref{fig:BERc2} and \ref{fig:BERd2} since for this case $t^{*} \approx 5.3s$.

\section{Conclusion}

Microfluidic channels with flow is a  very good candidate for molecular communication since the boundaries and flow reduce the ISI and increase the data rate, which are still considered as  open problems for many channel types. Therefore, the distribution of molecules in the channel and the impulse response  are necessary to determine the characteristics of the channel. Although there are many different attempts for the derivation of the impulse response of microfluidic channels, all consider some specific cases by reducing the advection-diffusion equation to the case for a 1-D channel, which is valid for uniform radial distribution. On the other hand, it may take some time to have  uniform radial distribution,  for smaller diffusion coefficient or higher channel radius. Until that time, the reduced 1-D model to solve axial distribution, hence that channel impulse response, is not applicable for these cases. We, therefore, firstly derive the radial distribution of the molecules inside the channel with respect to time, and accordingly determine the required time that the molecules reach the uniform radial state. Then, using the radial distribution and the time that molecules reach the uniform state, we derive the piecewise channel impulse response. Finally we verify the derived formulas of the  channel impulse response and radial distribution of molecules with Monte Carlo simulations.

\section{Acknowledgements}
This research was partially supported by the Scientific and Technical Research Council of Turkey (TUBITAK) under Grant number 116E916 and by TETAM: Telecommunications and Informatics Technologies Research Center at Bogazici University under grant number DPT-2007K120610. Fatih Din\c{c} would like to thank Professor Achim Kempf for insightful discussions on the spectral theory of the Laplacian operator.

\end{document}